\documentclass{article}
\usepackage[preprint]{log_2024}			% for preprint version

\usepackage{booktabs}						% professional-quality tables
\usepackage{multirow}						% tabular cells spanning multiple rows
\usepackage{amsfonts}						% blackboard math symbols
\usepackage{graphicx}						% figures
\usepackage{duckuments}						% sample images

\usepackage[numbers,compress,sort]{natbib}	% for numerical citations

\usepackage{amsmath,amssymb,amsfonts}
\usepackage{algorithmic}
\usepackage{graphicx}
\usepackage{textcomp}
\usepackage[american]{babel}
\usepackage{mathtools} % amsmath with fixes and additions
\usepackage{booktabs} % commands to create good-looking tables
\usepackage{tikz} % nice language for creating drawings and diagrams
\usepackage{algorithm}
\usepackage{algorithmic}
\usepackage{multirow}
\usepackage{colortbl}
\usepackage{bbding}
\usepackage{subcaption}
\usepackage{soul}
\usepackage{comment}
\usepackage{graphicx}
\usepackage{listings}
\usepackage{bm}
\usepackage{graphicx}
\usepackage[hyperpageref]{backref}
\usepackage{amsmath}
\usepackage{fnpos}
\usepackage{graphicx}
\usepackage{times}
\usepackage{latexsym}
\usepackage{adjustbox}
\usepackage{booktabs}
\usepackage{subfiles}
\usepackage{amsmath}
\usepackage{tabularx}
\usepackage{comment}
\usepackage{siunitx}
\usepackage{pdflscape}
\usepackage{multirow}
\usepackage{soul}
\usepackage{enumitem}
\usepackage{caption}
\usepackage{adjustbox}
\usepackage{sepfootnotes}
\usepackage{longtable}
\usepackage{xspace}
\usepackage{algorithm}
\usepackage{wrapfig}
\usepackage[lined,vlined,ruled,linesnumbered,algo2e]{algorithm2e}

\newcommand{\AT}{CrysAtom\xspace}

\newcommand{\Skip}{SkipAtom\xspace}

\newcommand{\Atom}{Atom2Vec\xspace}

\newcommand{\Rand}{Random\xspace}

\newcommand{\hlt}[1]{\textcolor{blue}{#1}}

\title[\AT: Distributed Representation of Atoms for  Crystal Property Prediction]{\AT: Distributed Representation of Atoms for  Crystal Property Prediction}

\author[Shrimon Mukherjee]{%
Shrimon Mukherjee\thanks{Equal contribution.}\\
Indian Association for the Cultivation of Science, India\\
\email{shrimonmukherjee@gmail.com}\And
Madhusudan Ghosh\footnotemark[1]\\
Indian Association for the Cultivation of Science, India\\
\email{madhusuda.iacs@gmail.com}\And
Partha Basuchowdhuri\\
Indian Association for the Cultivation of Science, India\\
\email{partha.basuchowdhuri@iacs.res.in}
}

\begin{document}

\maketitle

\begin{abstract}
Application of artificial intelligence (AI) has been ubiquitous in the growth of research in the areas of basic sciences. Frequent use of machine learning (ML) and deep learning (DL) based methodologies by researchers has resulted in significant advancements in the last decade. These techniques led to notable performance enhancements in different tasks such as protein structure prediction, drug-target binding affinity prediction, and molecular property prediction. In material science literature, it is well-known that crystalline materials exhibit topological structures. Such topological structures may be represented as graphs and utilization of graph neural network (GNN) based approaches could help encoding them into an augmented representation space. Primarily, such frameworks adopt supervised learning techniques targeted towards downstream property prediction tasks on the basis of electronic properties (formation energy, bandgap, total energy, etc.) and crystalline structures. Generally, such type of frameworks rely highly on the handcrafted atom feature representations along with the structural representations. In this paper, we propose an unsupervised framework namely, \AT, using untagged crystal data to generate dense vector representation of atoms, which can be utilized in existing GNN-based property predictor models to accurately predict important properties of crystals. Empirical results show that our dense representation embeds chemical properties of atoms and enhance the performance of the baseline property predictor models significantly.
\end{abstract}

\section{Introduction}
\begin{figure}[t]%
    \centering
    % \hspace{-1.5cm}
    {\includegraphics[width=0.65\textwidth]{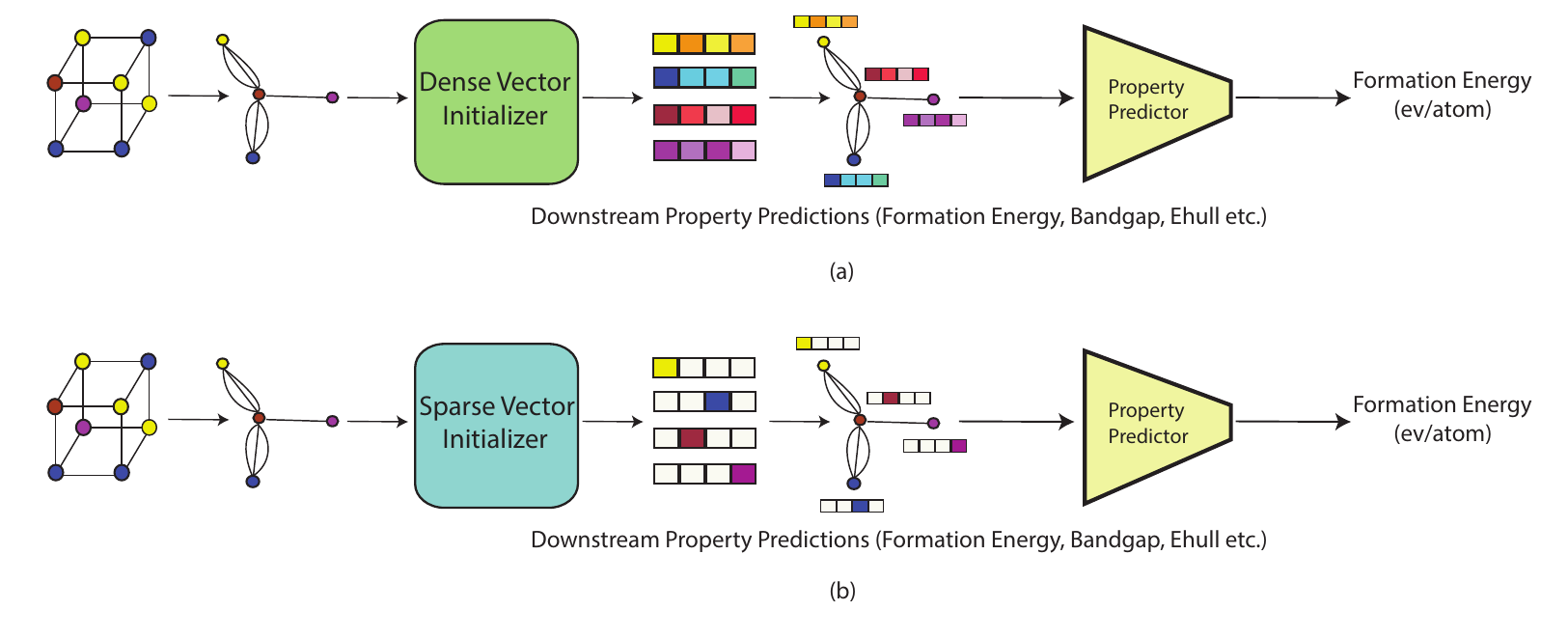}}
    {\caption{\small Workflow showing the downstream property prediction tasks with the (a) application of the distributed vector representation obtained from \AT framework, and (b) application of sparse atom vector representation.}
    \label{fig:atomgnn_intro}}
\end{figure}

In recent years, there has been a significant surge in applying machine learning (ML) algorithms across various disciplines, including material science and chemistry, where ML advancements are leveraged to address domain-specific challenges~\cite{jumper2021highly,artrith2021best,choudhary2022recent,mukherjee2022deepglstm}, such as property prediction, molecule generation, and discovery of key descriptors for CO2 activation~\cite{xie2018crystal,rengaraj2023two}. Despite the reliance on density functional theory (DFT) simulations in early material science works~\cite{das2022crysxpp}, their resource-intensive nature prompted a shift towards ML-based strategies to replace high-latency computational processes with efficient approximations~\cite{davies2016computational,antunes2022distributed}.
ML techniques depend on handcrafted features, whereas deep learning algorithms learn feature representations, mitigating the need for domain-expert intervention~\cite{das2022crysxpp}. In material science, crystal structures play a significant role for most of the downstream tasks~\cite{xie2018crystal}. Since majority of the crystals are available in nature as three-dimensional (3D) structures, they are initially transformed into graphs by preserving their periodic invariance~\cite{xie2018crystal}. Such graphs are used in graph neural network based frameworks for solving different downstream property prediction tasks~\cite{xie2018crystal,choudhary2021atomistic,yan2022periodic,pmlr-v202-lin23m}.\\ 
For the downstream tasks, atoms are commonly initialized using a one-hot sparse representation, leading to suboptimal performance~\cite{zhou2018learning,antunes2022distributed}. In contrast, distributed representations encapsulate richer semantic and structural information~\cite{mikolov2013efficient,pennington2014glove}. Atom2Vec~\cite{zhou2018learning} proposed a singular value decomposition (SVD) based distributed atom vector representation using handcrafted feature vectors, requiring domain knowledge. To alleviate this problem, SkipAtom~\cite{antunes2022distributed} introduced a skip-gram~\cite{pennington2014glove} based dense representation of the atoms by learning the required feature representation from the dataset. However, neither consider structural representation of the crystal materials, that can be harnessed by neural network-based models for improving dense representations. To mitigate this challenge, we investigate the feasibility of utilizing the graph structure information into the neural network framework towards generating distributed atom vector representations. On this note, we propose a novel auto-encoder-decoder based framework, namely \underline{Crys}tal \underline{Atom} Vector Extractor (\textbf{\AT}), to learn distributed representations of molecular atoms (shown in Figure~\ref{fig:atomgnn_intro}) by introducing a fusion mechanism by combining Self-Supervised Learning (SSL) and Unsupervised Learning (UL)-based techniques. The key distinction of our approach from existing work~\cite{das2023crysgnn} lies in the adoption of a novel fusion mechanism, characterized by a generalized SSL loss for pretraining task. Unlike previous methods, our SSL techniques are uniquely generalized, requiring no external information such as space groups. Furthermore, we utilize our proposed distributed vector representation for different downstream property prediction tasks and analyze its performance.
\\
\noindent{\textbf{Our Contributions. }}\\
% \vspace{-1.0mm}
1. To the best of our knowledge, we are the first to investigate the feasibility of applying an auto-encoder-decoder based graph neural network approach to obtain a domain-independent generic distributed representation of atoms.\\
2. We assess the quality of our distributed representation by comparing it with other existing state-of-the-art (SOTA) representations of atoms (\Atom and \Skip) in multiple property prediction tasks.\\
3. We use two popular benchmark materials datasets to show that our distributed representation of atoms obtained from \AT helps gain substantial improvement in performance for several property predictor models (CGCNN, ALIGNN) over their vanilla (from 5.21\% to 21.92\%), distilled and fine-tuned versions.\\ 
4. Moreover, the property-tagged dataset suffers from error bias, as it is theoretically derived from DFT. We successfully mitigate this issue using a small set of experimental data in the training setup.
\section{Related Work}
In recent times, data-driven approach specifically the graph neural network based frameworks~\cite{xie2018crystal,louis2020graph,wang2021compositionally} played a crucial role to conduct the property prediction task by utilizing the topological structures of the crystal materials.
Earlier studies~\cite{xie2018crystal,chen2019graph,chen2021exploring,choudhary2021atomistic,chen2021direct} did not comprise of the periodic invariance properties. Later, Matformer~\cite{yan2022periodic}, applied a periodic graph transformer based framework by employing both periodic invariance and periodic pattern encoding strategy. Similarly, PotNet~\cite{pmlr-v202-lin23m} used interatomic potentials for the property prediction tasks. Additionally, there are several works which employed UL~\cite{das2022crysxpp,das2023crysgnn} and SSL~\cite{stark20223d,magar2022crystal} strategies to apply the pretraining task followed by task-specific finetuning. All the discussed works employed sparse one-hot representations to initialize the node level feature representation. On this note, we propose our novel \AT framework towards generating distributed atom vector representation by incorporating graph structure as the necessary input representation into the encoder module and also by introducing generalized unsupervised contrastive loss,  which does not require any external information such as space group, as described in the work~\cite{das2023crysgnn}.
\section{Methodology}
% \begin{minipage}{0.6\textwidth}
\begin{table}
    \centering
    \small
    % \scalebox{0.7}{
    % \vspace{-2mm}
    \resizebox{0.45\columnwidth}{!}{\begin{tabular}{c|c}
    \hline
         {\bf Notation} & {\bf Terms}  \\
         \hline
         $u$ & Source node\\
         $v$ & Target node\\
         $V_{i}$ & Set of nodes present in the unit cell\\
         $n_{uv}$ & No. of edges between $u$ and $v$\\ 
         $E_{i}$ & A multiset of node pairs\\
         $\chi_{i}$ & Node features\\
         $F_{i}$ & Collection of edge weights\\
         $s$ & Length of one bond\\
         $r$ & Radius\\
         $\mathcal{M}$ & Cross-correlation matrix\\
         $READOUT$ & Global pooling function\\
         \Rand & Vector representations drawn from $\mathbb{R}^{n}$ using normal distribution \\
         $\mathcal{I}$ & Target cross-correlation matrix\\
         $\mathcal{L_{BT}}$ & Barlow Twins Loss\\
         $\mathcal{L_{FR}}$ & Node feature reconstruction loss\\
         $\mathcal{L_{CR}}$ & Local connectivity reconstruction loss\\
         \hline
    \end{tabular}}
    % }
    \caption{\small Important notations used in the paper.}
    \label{tab:notation}
    % \vspace{-5mm}
\end{table}
\begin{figure}[t]%
    \centering
    % \hspace{-1.5cm}
    {\includegraphics[width=0.50\textwidth]{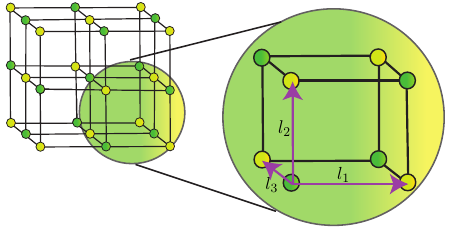}}
    {\caption{\small This illustration showcases a periodic crystal structure, featuring a point cloud of atoms arranged in repeating patterns. The image includes a magnified view of a unit cell, clearly delineating the lattice vectors \( L = [l_1, l_2, l_3] \), highlighting the fundamental building blocks of the crystal's geometric arrangement.}
    \label{fig:periodic_crystals}}
\end{figure}

\begin{figure}[t]%
    \centering
    % \vspace*{-1mm}
    {\includegraphics[width=0.65\textwidth]{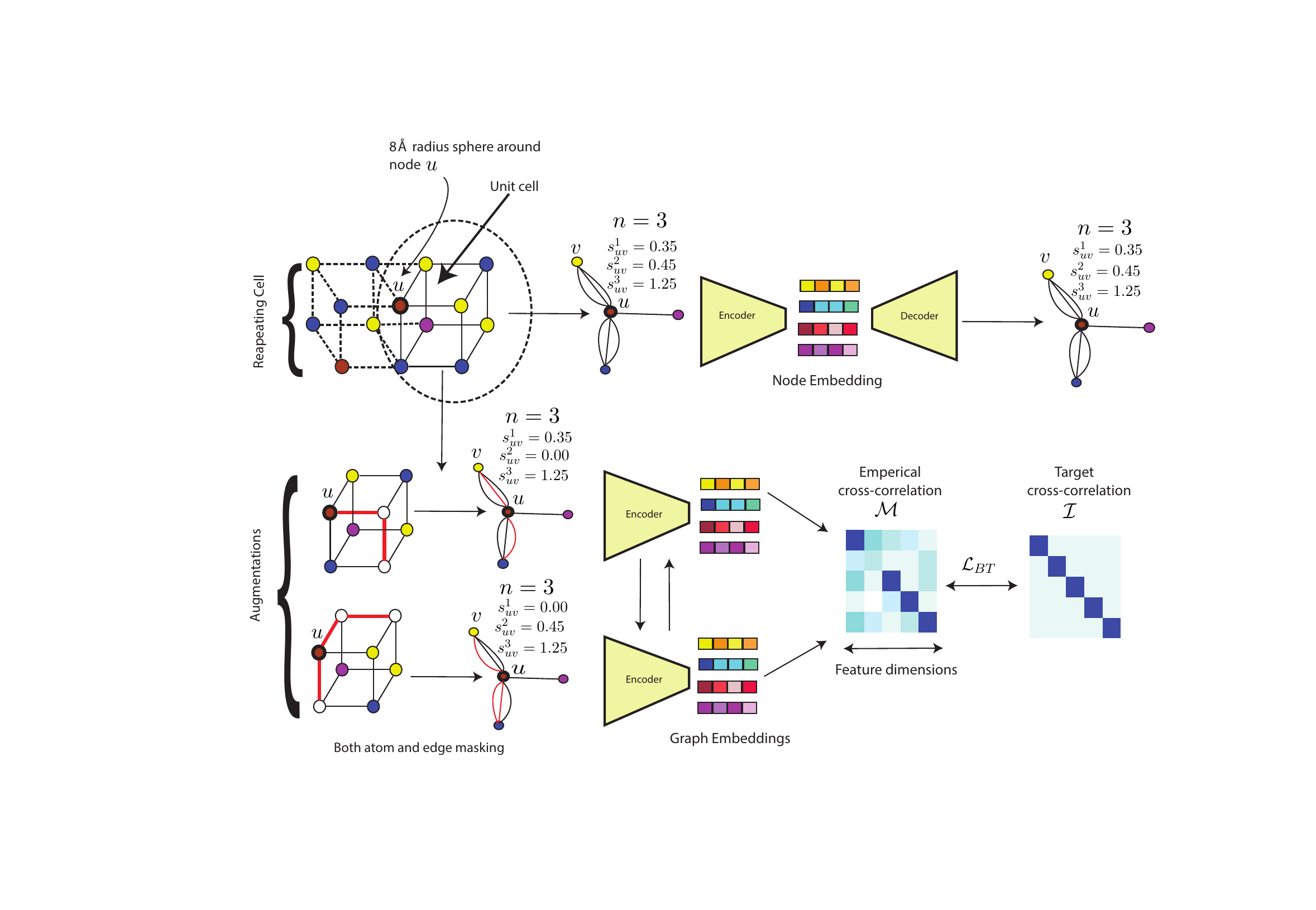}}
    {\caption{\small Schematic diagram illustrating the architecture of our proposed framework, \textbf{\AT}, highlighting the key components. The diagram encapsulates the novel mechanisms underlying \AT's functionality.}
    \label{fig:atomgnn}}
\end{figure}

% \end{minipage}
This section commences by outlining the general idea of our proposed neural framework, namely \AT, for generating dense vector representation of chemical atoms. Subsequently, a detailed discussion about the different components of our proposed framework follows. An overview of our proposed \AT framework has been shown in Figure~\ref{fig:atomgnn}. Table~\ref{tab:notation} summarizes important terms and corresponding notations used in this work.
\begin{figure}[t]%
    \centering
    % \hspace{-1.5cm}
    {\includegraphics[width=0.65\textwidth]{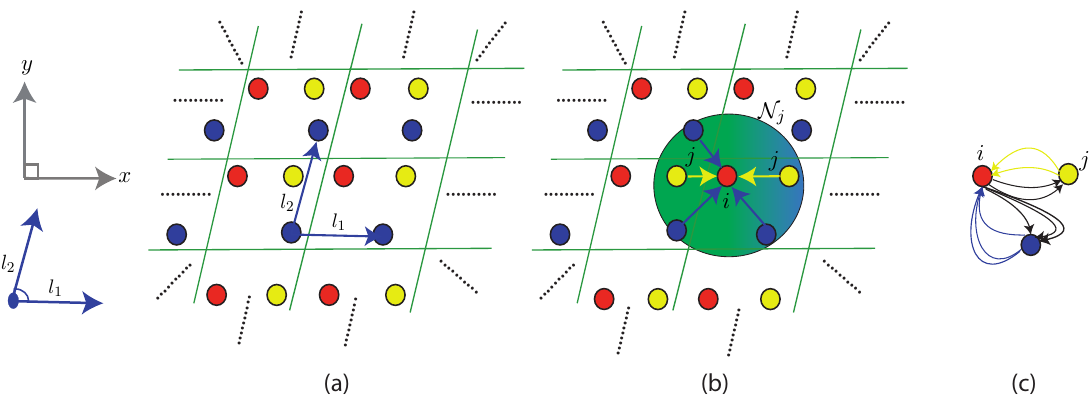}}
    {\caption{\small Illustration of the multigraph representation of a crystal. We use blue arrows, $l_1$ and $l_2$, for the a crystal structure in 2D. We use circles of different colors to represent different atoms and green lines to denote periodic boundaries. We use $\mathcal{N}_{i}$ as the neighborhood set of node $i$, and use yellow and blue arrows to specify the captured atomic interactions from the multigraph of the given crystal. (a) A crystal structure with periodic patterns $l_1$ and $l_2$ is shown in 2D. (b) Atomic interactions between the yellow nodes $j$ and the center red node $i$, captured by the multigraph of the crystal. (c) The corresponding multigraph. All the periodic duplicates of $j$ in the crystal are mapped to a single node $j$ in the multigraph.}
    \label{fig:crystal_multigraph}}
\end{figure}
\subsection{\AT}
In this work, we propose a novel encoder-decoder based neural framework, \AT, to generate dense vector representations\footnote{Upon acceptance of our work, we will release our atom vector representation.} of the chemical  atoms, which can be used to enhance the performance of SOTA downstream neural property predictor models present in the literature of material science~\cite{xie2018crystal, yan2022periodic, das2022crysxpp, das2023crysgnn}. To generate the dense vector representation, we first consider a collection of untagged crystal graphs $D_{ut}=\{G_{i}\}$ collected from the well-known materials database, which serves as input to our proposed \AT model ($f_{\theta}$). Similar to some of the state-of-the-art methods, popularly known for generating dense vector representations of atoms, such as \Skip and \Atom, we use untagged data for generating dense vectors with better generalization capabilities. Subsequently, it generates the dense vector representation of the chemical atoms by learning the intrinsic structural and chemical patterns from the input representation of the crystal graphs. Additionally, to establish the effectiveness of the vector representations generated by \AT, we conduct extensive theoretical and empirical analysis. For empirical analysis, we consider chemical property prediction, which is an important and challenging task in material 
science literature~\cite{das2022crysxpp,das2023crysgnn}.\\
\noindent\textbf{Crystal Graph Representation.}
As proposed by Xie et. al.~\cite{xie2018crystal}, we use crystal graph structures $D = \{G_{i} = (V_{i},E_{i},\chi_{i},F_{i})\}$ to represent crystalline materials. A crystal lattice is formed by repeating a unit cell in all three dimensions as shown in Figure~\ref{fig:periodic_crystals}. $G_{i}$ is an undirected weighted multigraph that represents unit cell of the crystal structure. $V_{i}$ denotes a set of nodes (atoms) present in the unit cell and $E_{i}=\{(u,v,n_{uv})\}$ denotes a set of triples, where each triple consists of three entries - a pair of nodes and the number of edges between them. It signifies that $v$ is an atom that appears $n$\textsubscript{$u$$v$} times in the \emph{nearby cells} around $u$. The \emph{nearby cells} are defined as the cells that are within a distance of radius $r$ from $u$. Therefore, $r$ is a hyper-parameter for this model and the set of \emph{nearby cells} may be expressed as, 
\begin{center}
   $\mathcal{N}_{r}(u)=\{v \in V_i \hspace{1mm}|\hspace{1mm}dist(u,v)\leq r \}$, 
\end{center}
 where $V_i$ is the set of atoms present in the crystal graph $G_i$ of crystal $i$ and $\mathcal{N}_{r}(u)$ is the set of \emph{nearby cells} for $u$ given $r$. In the later part of the paper, we have often used the terms nodes and atoms interchangeably. The pictorial representation of creating multigraph from crystal structure has been shown in Figure~\ref{fig:crystal_multigraph}. $\chi_{i}$ represents node features i.e., features which comprehensively represent the chemical properties of an atom, such as atomic volume, electron affinity, etc. Finally, $F_{i}$ represents a collection of edge weights between a pair of atoms in a crystal graph. In other words, $F_{i}=\{\{s^{n}\}_{(u,v)}\hspace{1mm}|\hspace{1mm}(u,v) \in E_{i}\}$ is a set that represents a collection of bond length values between each pair of nodes $(u,v)$ that are connected by an edge in $E_{i}$. The bond length is denoted as $s^{n}$, where $s$ is the length of one bond and $n$ is the number of bonds (hence, the number of edges) between $u$ and $v$. We consider bond length as a measure of distance from one atom to other atoms in close proximity. 
In the next section, we explain our proposed methodology for generating dense vector representation of atoms and subsequently analyze its effectiveness in property predictor models.
\subsection{Atom vector formation}
In this part, we discuss the architecture of the proposed CrysAtom model, as shown Figure~\ref{fig:atomgnn}. It consists of an auto-encoder with an SSL framework that leverages the correlations in the input to learn robust and generalizable dense vector representation of atoms.\\
\noindent\textbf{Encoder.}
We develop our auto-encoder module by employing crystal graph convolutional neural network (CGCNN)~\cite{xie2018crystal}. We utilize CGCNN to encode the chemical and structural information of a crystal graph $G$. It encodes information of the $l$-hop neighborhood for each node by 
applying following equations:
\begin{equation}
% \tag{1}
\begin{aligned}
&z_{(u,v)_{n}}^{l-1} = x_{u}^{l-1} \oplus x_{v}^{l-1} \oplus s^{n}_{(u,v)}
\end{aligned}\label{eq:enc}
\end{equation}

\begin{equation}
\centering
% \hspace{-4.5mm}
\begin{aligned}
&x_{u}^{l} = x_{u}^{l-1} + \sum_{v,n} \sigma(z^{l-1}_{(u,v)_{n}}\Theta_{c}^{l-1} + b_{c}^{l-1}) \odot g(z_{(u,v)_{n}}^{l-1}\Theta_{s}^{l-1} + b_{s}^{l-1}) \notag
\end{aligned}
\end{equation}
where, $l$ is the number of CGCNN layers, $x_{u}^{l-1}$ denotes the embedding of node $u$ by aggregating the $l-1$ hop neighborhood information. The embedding of node $u$ is initialized to a transformed node feature vector, i.e., it is a function of the atom $u$'s chemical features such as ${x}_{u}^{0} = \chi_{u}\Theta_{\chi}$ where $\Theta_{\chi}$ is the list of trainable parameters of the transformation network and $\chi_{u}$ is the input node feature vector. $s_{(u,v)}^{n} \in F_{u}$ represents the overall bond length between atoms $u$ and $v$. The operator $\oplus$ denotes concatenation and $\odot$ denotes element-wise multiplication. 
Here, $\Theta_{c}^{l-1}$, $\Theta_{s}^{l-1}$, $b_{c}^{l-1}$, $b_{s}^{l-1}$ are the convolution matrix, self-weight matrix, convolution bias, self-bias of $(l-1)$th layer convolution, respectively. $\sigma$ is a nonlinear transformation function, used to generate a real value in [0,1] indicating the edge importance and $g$ is a feed-forward network. Finally, we collect local information at each node after aggregating the information from the neighborhood ($x_{u}^{l}$). We denote the set of trainable parameters for this encoder as $\Theta_{e}$ for future reference.\\
\noindent\textbf{Decoder.}
% \hspace{0.15cm}
\label{sec:decoder}
The encoder encodes the chemical properties of an atom into a latent vector space $x$ by learning the structural and chemical information. Subsequently, the decoder tries to decode the vectors from $x$, thereby enhancing the encoding capability of the encoder.
As mentioned earlier, the crystal properties depend on the local chemical environment and the overall conformation of the repeating crystal cell structure. Therefore, we build our decoder framework to reconstruct two important features to capture the properties of local chemical environment: (a) the node features, which are the chemical properties of individual atoms, and (b) the local connectivity, which is the relative position of the nodes with respect to their local neighbors. We employ the node feature reconstruction strategy by computing the following equations,
\begin{equation}
\begin{aligned}
&\hat{\chi_{u}} = \Theta_{x}^{T}x_{u}^{l} + b_x 
\end{aligned}
 \label{eq:l1}
\end{equation}
\begin{equation}
\begin{aligned}
&\mathcal{L_{FR}} = -\chi_u \cdot \log(\hat{\chi_u}) - (1 - \chi_u) \cdot \log(1 - \hat{\chi_u})
\end{aligned}
 \label{eq:fr}
\end{equation}
% A combined transformed embedding of nodes $u$ and $v$ is denoted by $x_{uv}$. The length of the bonds is given by a real number that is generated by a feed-forward network $\beta_{s}$.\\
where $\Theta_x$, $b_x$ and $\mathcal{L_{FR}}$ are the trainable weights, biases and the feature reconstruction loss, respectively.\\
Furthermore, we reconstruct the global topological information to generate the connectivity and the periodicity information of the crystal structures by employing a bilinear transformation strategy, with the help of the following equations,
\begin{equation}
\begin{aligned}
&x_{uv}^{l} = \sigma(x_{u}^{l}\Theta_{bl}^{T}x_{v}^{l} + b_{bl})
\end{aligned}
\label{eq:g1}
\end{equation}

\begin{equation}
\begin{aligned}
&\mathcal{L_{CR}} = \arg \max_n \frac{e^{\beta_{k}(x_{uv}^{l},n)}}{\sum_{n}e^{\beta_{k}(x_{uv}^{l},n)}}
\end{aligned}
\label{eq:CR}
\end{equation}
where \(\Theta_{bl}\), \(b_{bl}\) and $\mathcal{L_{CR}}$ are the trainable weights and biases and connection reconstruction loss, respectively. The function \(\sigma\) is an activation function that maps the input to a value between 0 and 1. The output representation from the initial bilinear transformation layer (\(\Theta_{bl}\) and \(b_{bl}\)) is passed to another linear transformation layer followed by a softmax activation. The term \(\beta_{k}\) represents a feed-forward neural network with \(k\) layers, which generates a logit vector of length \(n\) (as mentioned previously, $n$ is the number of edges between two atoms) with the help of a softmax function.
% Equations~\ref{eq:g2},~\ref{eq:l1} show that both global and local information are closely related to each other. 
We denote the set of trainable parameters for this decoder as $\Theta_{d}$ for future reference.
\DontPrintSemicolon
\begin{algorithm}[t]
        \algsetup{linenosize=\footnotesize}
        \footnotesize
	\SetKwInOut{Input}{Input}
	\SetKwInOut{Output}{Output}
	\SetCommentSty{textrm}
	\SetKwComment{Comment}{$\triangleright$}{}
	\Input{$D = \{G_{i} = (V_{i},E_{i},\chi_{i},F_{i})\}$ dataset used for creating the chemical atom vector }
	\Output{ Generalized dense atom vector representations ($A$)}
        % \hline
        \hrule
	\BlankLine
        \Begin{
        Initialize $\Theta_{e}$ \hlt{$\triangleright$ Parameters of the encoder}
        
        Initialize $\Theta_{d}$ \hlt{$\triangleright$ Parameters of the decoder}
        
        $\mathcal{L_{FR}}$ \hlt{$\triangleright$ Mean squared error loss (Equation~\ref{eq:fr})\;}
        
        $\mathcal{L_{CR}}$ \hlt{$\triangleright$ Connection reconstruction loss (Equation~\ref{eq:CR})}\;
        
        $\mathcal{L_{BT}}$ \hlt{$\triangleright$ Barlow Twins loss (Equation~\ref{eq:barlowloss_new})\;}

        $A$ \hlt{$\triangleright$ Set of atom vectors\;}

        $S$ \hlt{$\triangleright$ Set of common atoms in one batch\;}

        $N$ \hlt{$\triangleright$ Number of epochs\;} 
        
        % $G_{j} = $ batched graph\; 
        $G_j$ \hlt{$\triangleright$ Batch graph (collection of 128 crystal structures)\;}
        
        $D$ \hlt{$\triangleright$ Graph data\;}

        \For{$i \gets 1$ to $N$}{
            \For{$G_{j} \in D$}{
                \hlt{$\triangleright$ $j$-th batch with randomly selected 128 crystals } \\
                $H_{emd}$ = $f_{e}$($G_j$, $\Theta_e$)\;\\ \hlt{$\triangleright$ $H_{emd}$: hidden representation of $G_j$}

                $A_{Femd},$ $Adj_e$ = $f_d$($H_{emd}$, $\Theta_d$)\;
                \\ \hlt{$\triangleright$ $AF_{emd}$: output of the decoder, $Adj_{e}$: constructed adjacency matrix}
                %\Comment{$AF_{emd}$ is the output of the decoder, $Adj_{e}$ is the constructed adjacency matrix.}

                $G_1, G_2\gets G_{j}$\;\\ \hlt{$\triangleright$ $G_1$, $G_2$: augmented representations of $G_j$}%\Comment{$G_1$, $G_2$ are augmented representations of $G_j$}

                $\mathcal{X}^{G_1}, \mathcal{X}^{G_2}=f_{e}(G_1, G_2)$\; \\ \hlt{$\triangleright$ $\mathcal{X}^{G_1}, \mathcal{X}^{G_2}$: augmented embeddings of $G_1, G_2$}

                $\mathcal{L}_{j} = \alpha\mathcal{L_{FR} + \beta\mathcal{L_{CR}}} + \gamma\mathcal{L_{BT}}$\;

                $\Theta= \Theta_{e} \oplus \Theta_{d}$\;

                $\Theta = \Theta\ -\ \alpha \nabla \mathcal{L}$\;

                $A_j \gets f_{Unbatch}(H_{emd})$\; \\
                \hlt{$\triangleright$ $f_{Unbatch}$ generates atom representations}

                $\forall$ $s \in S$, $S_{j}^{s}$ = $0$, $count_s$ = 0\; \\
                \hlt{$\triangleright$ $S^{s}_{j}$: cumulative representation of atom $s$ in $A_j$}

                $\forall$ $A_{j}^{s} \in A_j$, $count_s++$, $S_{j}^{s} = S_{j}^{s} + A_{j}^{s}$\;

                $\forall$ $s \in S$, $A_{j}^{s} = \frac{S_{j}^{s}}{count_s}$\; \\ \hlt{$\triangleright$ Final representation of atom $s$ for batch $j$}

                 \eIf{$\mathcal{L}_j\leq \mathcal{L}_{j-1}$}
                 {
                    $A_{j} = A_{j}$\;
                 }
                 {
                    $A_{j} = A_{j-1}$\;
                 }
            }
		}
        return $A$\;
    }
	\caption{Training procedure and chemical vector extraction}
        \label{algo:atomgnn_vec}
\end{algorithm}
\\
\noindent\textbf{SSL framework.} 
% \hspace{0.15cm}
\label{sec:SSL}
% In this section, we discuss our SSL framework that is compatible with a UL-based framework. 
Here, the SSL framework uses correlation between the actual input graph and its augmented versions to learn robust and generalized representation from the unlabeled data~\cite{magar2022crystal}. It provides an additional boost in terms of performance. We employ two types of augmentation techniques such as atom masking and edge masking. Atom masking randomly masks 10\% of the atoms in the crystal, while edge masking randomly masks 10\% of the edge features between adjacent atoms.
\begin{equation}
    \begin{aligned}
        \mathcal{X}_{G} = READOUT({x_{u}^{l})}
    \end{aligned}
    \label{eq:SSL1}
\end{equation}
where READOUT interpretes to a global pooling function. Augmented graph representations are  generated from the same crystalline structures. These augmented representations are used towards identity matrix formation. Given the problem formulation, we identified Barlow Twins~\cite{zbontar2021barlow} (BT) loss as a suitable loss function to reconstruct the graph representation of the crystals. Hereby, we utilize Barlow Twins loss function, which is based on the redundancy reduction principle by H. Barlow~\cite{barlow1961possible,barlow2001redundancy}. We apply this loss function to the cross-correlation matrix that is formed from the embeddings produced by the encoder module of the auto-encoder. 
% We apply the augmented instances to the crystal graphs for generating dense vector representation of atoms.
\begin{equation}
 \begin{aligned}
    \mathcal{L_{BT}} \triangleq \underbrace{\sum_{i} (1-\mathcal{M}_{ii}^{2})}_\text{invariance term} + ~~\lambda \underbrace{\sum_{i}\sum_{j \neq i} \mathcal{M}_{ij}^{2}}_\text{redundancy reduction term}
    \end{aligned}
    \label{eq:barlowloss_new}
\end{equation}

\begin{equation}
    \begin{aligned}
    \mathcal{M}_{ij} \triangleq \frac{\sum_{b}\mathcal{X}_{b,i}^{G_{1}}\mathcal{X}_{b,j}^{G_{2}}}{\sqrt{( \mathcal{X}_{b,i}^{G_{1}})^{2}}\sqrt{(\mathcal{X}_{b,j}^{G_{2}})^{2}}}
    \end{aligned}
    \label{eq:barlowloss2}
\end{equation}
Equation~\ref{eq:barlowloss_new} describes the Barlow Twins loss function, which is used in our SSL block. It considers the cross-correlation matrix $\mathcal{M}$ of embeddings from two augmented instances, which is computed by Equation~\ref{eq:barlowloss2}. The parameter $\lambda$ (set to 0.0051 in the original paper~\cite{zbontar2021barlow}) is a positive constant that balances the first and second terms of the loss function. In this study, we apply Equations~\ref{eq:enc} and \ref{eq:SSL1}, to obtain the two augmented embeddings, namely $\mathcal{X}^{G_1}$ and $\mathcal{X}^{G_2}$. Furthermore, we utilize Equation~\ref{eq:barlowloss2} to derive the cross-correlation matrix, wherein $b$ represents the batch index and $i,j$ indicate the vector dimension of the projected output. Overall, the deep auto-encoder architecture is trained in an end-to-end fashion to optimize the loss function ($\mathcal{L}_{train}$) as shown in Equation~\ref{eq:loss_final}.
\begin{equation}
    % \small
    \begin{aligned}
        \mathcal{L}_{train} = \underbrace{\alpha\mathcal{L_{FR}} + \beta\mathcal{L_{CR}}}_\text{loss for UL} + 
\underbrace{\gamma\mathcal{L_{BT}}}_\text{loss for SSL}
    \end{aligned}
    \label{eq:loss_final}
\end{equation} $\mathcal{L_{FR}}$, $\mathcal{L_{CR}}$ are the reconstruction losses for node features and local connectivity, respectively. $\mathcal{L_{BT}}$ is the Barlow Twins loss and $\alpha, \beta, \gamma$ are the weighting coefficients for each loss. The sequential steps for the training process and extraction of dense vector representation are stated in Algorithm~\ref{algo:atomgnn_vec}. The number of parameters used by \AT is 5.5 MB and the running time of each epoch of the training (using Algorithm~\ref{algo:atomgnn_vec}) is approximately 30 minutes.

\noindent\textbf{Atom Feature Vector Extraction.}
The atom vector extraction strategy in Algorithm~\ref{algo:atomgnn_vec} focuses on deriving atom feature vectors from the latent feature representations generated by the encoder module during each training epoch. For a batch graph \( G_j \), the hidden representation \( H_{emd} \) is obtained using the encoder function \( f_{e} \):
   \[
   H_{emd} = f_{e}(G_j, \Theta_e)
   \]
Subsequently, the function \( f_{Unbatch} \) is applied to \( H_{emd} \) to generate individual atom representation \( A_j \).
   \[
   A_j = f_{Unbatch}(H_{emd})
   \]
Here, \( f_{Unbatch} \) maps the latent feature vectors \( H_{emd} \) to their corresponding atom feature vectors. If \( H_{emd} \) is a hidden representation of size \( n \times d \), where \( n \) is the number of nodes (atoms) and \( d \) is the dimensionality of the feature vector, \( f_{Unbatch} \) effectively extracts the important features to generate the distributed vector representation for each atom.

For each atom type \( s \) in the set of common atoms \( S \), the proposed algorithm computes a cumulative representation \( S_{j}^{s} \) and calculates \( {count}_s \) as,
   \[
   S_{j}^{s} = \sum_{i=1}^{n} \mathbb{1}_{\{A_{j}^{i} = s\}} \cdot A_{j}^{i}
   \]
   \[
   count_s = \sum_{i=1}^{n} \mathbb{1}_{\{A_{j}^{i} = s\}}
   \]
   Here, \( \mathbb{1}_{\{A_{j}^{i} = s\}} \) is an indicator function that is 1 if the atom \( A_{j}^{i} \) is of type \( s \) and 0 otherwise. Lastly, the final representation for each atom type \( s \), in the batch \( j \), is computed by averaging the cumulative representations.
   \[
   A_{j}^{s} = \frac{\sum_{i=1}^{n} \mathbb{1}_{\{A_{j}^{i} = s\}} \cdot A_{j}^{i}}{\sum_{i=1}^{n} \mathbb{1}_{\{A_{j}^{i} = s\}}}
   \] In this way, our algorithm ensures that the atom vectors are consistently updated and refined through the training epochs, ultimately storing the generalized dense atom vector representations in \( A \).
\subsection{Downstream Property Prediction Task}
The objective of this study is to integrate the proposed atom feature vectors into a SOTA property predictor model to enhance the performance of the downstream task. 
% Our dense feature vector of atoms is obtained through the implementation of our proposed \textbf{\AT} (see detail description
% of Algorithm in Appendix). 
The following steps are used for applying the dense vector representations to the downstream property prediction tasks. We extract the atom vectors from our novel \AT framework. Subsequently, we train SOTA property predictor model ($P_{\psi}$) by providing property-tagged training data $D_{t} = \{G_{i},y_{i}\}$ as well as the generated feature vector as initial node feature representation.  Here, we consider CGCNN~\cite{xie2018crystal}, CrysXPP~\cite{das2022crysxpp}, ALIGNN~\cite{choudhary2021atomistic}, Matformer~\cite{yan2022periodic} and PotNet~\cite{pmlr-v202-lin23m} as the baseline property predictors due to its SOTA performance in property prediction tasks. Training setup and hyper-parameter details are stated in Table~\ref{tab:parameters}.
\begin{table}[ht]
    \centering
    \small
    \resizebox{0.5\textwidth}{!}{%
    \begin{tabular}{| c | c |}
        \hline
        \textbf{Training Setup/} & {}
        \\
        \textbf{Hyper-parameter details/} & \textbf{Value} \\
        \textbf{Computational Resources} & {}\\
        \midrule
        $r$ & 8 \\
        Convolution Layers & Five convolution layers~\cite{xie2018crystal} \\
        Epochs & 110 \\
        Optimizer & Adam~\cite{kingma2014adam} \\
        Learning Rate & 0.03 \\
        Embedding Dimension & 200 (can be 50 and 100-dimensional dense vectors) \\
        Batch Size & 128 \\
        Weightage for $\alpha$ & 0.25 \\
        Weightage for $\beta$ & 0.25 \\
        Weightage for $\gamma$ & 0.50 (for convex sum) \\
        \midrule
        \midrule
        Optimizer & Adam~\cite{kingma2014adam}\\
        Epochs & 1000\\
        Learning Rate & Default learning rates used in the vanilla versions\\
        Batch Size & 64\\
        Seed Value & 123\\
        Train, Valid, Test Splits & 80\%, 10\%, 10\%\\
        \midrule
        \midrule
        Implementation Framework & PyTorch\\
        Computational Resources & one NVIDIA A6000 48GB GPU and one NVIDIA A100 80GB GPU\\
        \hline
    \end{tabular}%
    }
    \caption{\small Summary of hyper-parameter details in \AT (top), \AT variant of all SOTA models (middle) and computational resources (bottom) used in this work.}
    \label{tab:parameters}
\end{table}

\section{Results}
In this section, we first describe the details of our dataset used in our experiments, and then we follow up with the research questions in context to the task of atom vector generation, and the analytical discussion towards addressing those research questions.
% \begin{table}
%   \centering
%     % \caption{Datasets Details}

%   \small
%     \setlength{\tabcolsep}{2 pt}
%     \scalebox{0.68}{
%     \begin{tabular}{c | c | c | c | c |}
%     \toprule
%     Task & Datasets & Graph Num. & Structural Info.
%     \midrule
%      \multirow{2}{*}{\shortstack{Pre-training}}
%      & OQMD & 661K & \checkmark  & x\\
%      & MP & 139K & \checkmark  & x\\
%      \midrule
%     \multirow{2}{*}{\shortstack{Property \\ Prediction}} & MP 2018.6.1 & 69K & \checkmark  & 2 \\
%      & JARVIS-DFT & 55K & \checkmark  & 5\\
%      %& OQMD-EXP & 1.5K & \checkmark  & 1 & Experimental\\
%     \bottomrule
%   \end{tabular}
%   }
%   \label{tbl-dataset}
% \end{table}

\begin{table}[ht]
  \centering
  \small
    \setlength{\tabcolsep}{2 pt}
    \resizebox{0.75\textwidth}{!}{
    \begin{tabular}{|c | c | c | c | c | c|}
    \toprule
    Task & Datasets & Graph Num. & Structural Info.  & Properties Count & Data Type\\
    \midrule
     \multirow{1}{*}{\shortstack{Dense Vector Creation}}
     & MP & 139K & \checkmark  & $\times$ & DFT  Calculated\\
     \midrule
    \multirow{3}{*}{\shortstack{Property \\ Prediction}} & MP 2018.6.1 & 69K & \checkmark  & 4 & DFT Calculated\\
     & JARVIS-DFT & 55K & \checkmark  & 7 & DFT Calculated\\
     & OQMD-EXP & 1.5K & \checkmark  & 1 & Experimental\\
    \bottomrule
  \end{tabular}
  }
  \caption{\small Datasets used for both dense vector creation and downstream tasks.}
  \label{tab:dataset_vector}
\end{table}
\begin{table}
  \centering
  \small
    \setlength{\tabcolsep}{10 pt}
  \resizebox{0.45\textwidth}{!}{\begin{tabular}{|ccc|}
    \toprule
    % \toprul
    % \multicolumn{2}{c}{Part}                   \\
    % \cmidrule(r){1-2}
    Property & Unit & Data-size\\
    \midrule
     Formation Energy & $eV/(atom)$ &  69239 \\
     Bandgap (OPT)    & $eV$ & 69239  \\
     Bulk Modulus (Kv)    & GPa & 5450 \\
     Shear Modulus (Gv)    & GPa & 5450 \\
    \midrule
    \midrule
    Formation Energy & $eV/(atom)$ & 55723 \\
    Bandgap (OPT)    & $eV$ & 55723 \\
    Total\_Energy    & $eV/(atom)$ & 55723\\
    Ehull    & $eV$ & 55371\\
    Bandgap (MBJ)    & $eV$ & 18172 \\
    
    Bulk Modulus (Kv)    & GPa & 19680\\
    Shear Modulus (Gv)    & GPa & 19680\\
     \bottomrule
  \end{tabular}}
  \caption{\small Summary of different crystal properties in Materials Project (top) and JARVIS-DFT (bottom) datasets.}
   \label{tbl-downstream}
\end{table}
\subsection{Datasets}
\label{datasets}
We use 139K unlabeled crystal graphs from the Materials Project (MP)\footnote{\url{https://materialsproject.org/}\label{mp}} to obtain the required dense vector representation of atoms. For our downstream property prediction, as suggested by the Yan et. al~\cite{yan2022periodic}, we consider the datasets MP 2018.6.1$^{\ref{mp}}$ and JARVIS-DFT 2021.8.12\footnote{\url{https://jarvis.nist.gov/}}~\cite{choudhary2020joint}, to investigate the chemical rationality of our proposed vector representations. 
% The MP dataset is a subset of the dataset used for our vector preparation, while JARVIS-DFT is a separate dataset that is not seen during the creation of the vectors. 
MP 2018.6.1 contains 69,239 materials with four properties, formation energy, bandgap (OPT), bulk modulus (Kv) and shear modulus (Gv), whereas the JARVIS-DFT dataset contains 55,723 materials with seven properties such as formation energy, bandgap (OPT), total energy, ehull, bandgap (MBJ), bulk modulus (Kv) and shear modulus (Gv). Details of each of these datasets are given in Table~\ref{tab:dataset_vector} and details of this properties is provided in Table~\ref{tbl-downstream}. Here, all these properties in Materials Project and JARVIS-DFT datasets are based on DFT calculations of crystal. To investigate how our dense vector representation mitigates DFT errors, we take a small dataset OQMD-EXP~\cite{kirklin2015open} containing 1,500 materials, consisting of experimental data for formation energy.
\begin{table*}[ht]
% \hspace{-9.9mm}
 \centering
  \small
    \setlength{\tabcolsep}{9 pt}
    \resizebox{1.0\textwidth}{!}{
\begin{tabular}{|c | c c | c c| c c | c c | c c|}
         \toprule
         Property &  CGCNN & CGCNN & CrysXPP & CrysXPP & ALIGNN & ALIGNN & Matformer & Matformer & PotNet & PotNet \\
         & & (\AT) & & (\AT) & & (\AT) & & (\AT) & & (\AT) \\
         \midrule
         \midrule
         Formation Energy & 0.039 & \textbf{0.028} & 0.041 & \textbf{0.030} & 0.026 & \textbf{0.023} & 0.021 & \textbf{0.019} & 0.019 & \textbf{0.018} \\
        Bandgap (OPT)  & 0.388 & \textbf{0.270} & 0.347 & \textbf{0.262} & 0.271 & \textbf{0.252} & 0.211 & \textbf{0.206} & 0.204 & \textbf{0.193} \\
        Bulk Modulus (Kv) & 0.054 & \textbf{0.050} & 0.080 & \textbf{0.048} & 0.051 & \textbf{0.043} & 0.043 & \textbf{0.040} & 0.040 & \textbf{0.038} \\
        Shear Modulus (Gv) &  0.087 & \textbf{0.082} & 0.105 & \textbf{0.082} & 0.078 & \textbf{0.072} & 0.073 & \textbf{0.071} &  0.065 & \textbf{0.064} \\
        \midrule
        \midrule
         Formation Energy & 0.063 & \textbf{0.040} & 0.062 & \textbf{0.041} & 0.033 & \textbf{0.031} & 0.033 & \textbf{0.030} & 0.029 & \textbf{0.028}\\
         Bandgap (OPT) & 0.200 & \textbf{0.143} & 0.190 & \textbf{0.142} & 0.142 & \textbf{0.133} & 0.137 & \textbf{0.135} & 0.127 & \textbf{0.120} \\
         Total Energy & 0.078 & \textbf{0.043} & 0.072 & \textbf{0.044} & 0.037 & \textbf{0.035} & 0.035 & \textbf{0.031} & 0.032 & \textbf{0.029} \\
         Ehull & 0.170 & \textbf{0.124} & 0.139 & \textbf{0.121} & 0.076 & \textbf{0.066} & 0.064 & \textbf{0.057}  & 0.055 & \textbf{0.049} \\
         Bandgap (MBJ) & 0.410 & \textbf{0.333} &  0.378 & \textbf{0.348} & 0.310 & \textbf{0.280} & 0.300 & \textbf{0.290} & 0.270 & \textbf{0.240} \\
         Bulk Modulus (Kv)  & 14.47 & \textbf{12.37} & 13.61 & \textbf{13.10} & 10.40 & \textbf{10.19} & 11.21 & \textbf{10.85} & 10.11 & \textbf{9.98} \\
         Shear Modulus (Gv) & 11.75 & \textbf{10.45} & 11.20 & \textbf{10.44} & 9.86 & \textbf{9.39} & 10.76 & \textbf{9.85} & 9.23 & \textbf{9.13} \\ 
        \bottomrule     
    \end{tabular}}
    \caption{\small Summary of the results (MAE) of different properties in Materials Project (top) and JARVIS-DFT
(bottom). Model M is the vanilla variant of a SOTA model and M(\AT) is a variant of the SOTA model with \AT dense vectors as input. The best performance has been highlighted in \textbf{bold}.}
\label{tab:main_result}
\end{table*}

\begin{table}
  \centering
  \small
    \setlength{\tabcolsep}{6 pt}
    % \scalebox{0.75}
    % {
    \resizebox{0.55\textwidth}{!}{
      \begin{tabular}{|c |c|c c c c|}
        \toprule
        % Property & CGCNN\textsubscript{\AT} & CGCNN\textsubscript{DT} & CrysGNN\textsubscript{FT} & CrysXPP & PT-GNN \\
        Property & CGCNN & Distilled & Fine-tuned & CrysXPP & Pre-trained\\
        & (\AT) & CGCNN & CrysGNN & & GNN\\
        % & &  & & & -GNN\\
        % &  &  &   \\
        \midrule
        Formation Energy  & \textbf{0.040} (\colorbox{green!20}{-14.9}) & \ul{0.047} & 0.056 & 0.062 &  0.764 \\
        Bandgap (OPT)  & \textbf{0.143} (\colorbox{green!20}{-10.6})   &  \ul{0.160} & 0.183 & 0.190 &  0.688 \\
        Total Energy   & \textbf{0.043} (\colorbox{green!20}{-18.9})  &  \ul{0.053} & 0.069 & 0.072 &  1.451 \\
        Ehull & \ul{0.124} (\colorbox{red!20}{+2.4}) &  \textbf{0.121} & 0.130 & 0.139 &  1.112 \\
        Bandgap (MBJ) & \textbf{0.333} (\colorbox{green!20}{-2.1}) &  \ul{0.340} & 0.371 & 0.378 &  1.493 \\
        Bulk Modulus (Kv) & \ul{12.37} (\colorbox{red!20}{+0.5}) & \textbf{12.31}  & 13.42 & 13.61 &  20.34 \\
        Shear Modulus (Gv) & \textbf{10.45} (\colorbox{green!20}{-3.9}) & \ul{10.87}  & 11.07 & 11.20 &  16.51 \\
        \bottomrule
      \end{tabular} 
      }
   \caption{\small Comparison of prediction performance (MAE) for the seven properties in JARVIS-DFT between \AT version of the CGCNN, Distilled CGCNN and other SOTA pre-trained/fine-tuned models. The best results have been shown in \textbf{bold} and the second best results have been \ul{underlined}. Percentage of decrease in MAE for \AT, with respect to the best performing model from the rest, has been mentioned within the bracket.}
  \label{tbl-distill-finetune}
\end{table}
\subsection{Research Questions}
We pose a few important research questions (RQs), which are central to our research work.
% As discussed earlier that initialization of atom information using sparse vector representation cannot produce a state-of-the-art (SOTA) performance in the neural downstream property prediction task in  material science literature. On that note, we investigate our first research question,
% \vspace{-4.5mm}
% of the existing neural property predictors? 
% aids the existing neural property predictor to achieve SOTA performance in comparison to the sparse vector representation as well as the existing vector representation?
% \vspace{-7.9mm}
%\para{RQ-2} To what extent does our dense representation of crystal atom improve the performance of the existing property predictor compared to distilled, pre-trained, and fine-tuned frameworks?

\noindent\textbf{RQ-1: Effectiveness of \AT for downstream property prediction task.}
How effectively does \AT aid the existing neural property predictors to achieve
SOTA performance? Furthermore, to what extent does \AT improve the performance of the distilled, pre-trained, and fine-tuned versions of the neural property predictors?

\noindent\textbf{RQ-2: Robust dense vector representation.} How does our dense vector representation \AT fare when compared to the popular vector representations for atoms, such as, \Atom or \Skip?

\noindent\textbf{RQ-3: Removing DFT bias.} How effectively does our proposed crystal atom vector representation mitigate DFT error bias, leading to SOTA results in existing neural property predictors?

\noindent\textbf{RQ-4: Preserving periodic properties.} 
How well does \AT capture the periodic properties and the chemical significance of the atom? Does it also account for the elements displaying aberrant behavior with respect to its position in the periodic table, such as, Hydrogen (H), Helium (He)?
\subsection{Discussions}
\noindent\textbf{Downstream Task Analysis.}
In relation to RQ-1, we compare five different SOTA frameworks for crystal property prediction such as CGCNN~\cite{xie2018crystal}, CrysXPP~\cite{das2022crysxpp}, ALIGNN~\cite{choudhary2021atomistic}, Matformer~\cite{yan2022periodic} and PotNet~\cite{pmlr-v202-lin23m}. To train these methods for property prediction, we use the 200-dimensional dense vectors obtained for each atom using Algorithm~\ref{algo:atomgnn_vec}. These vectors serve as input atom features, which are initialized as non-trainable node features. For each property, we trained on 80\%, validated on 10\% and evaluated on 10\% of the data.\\
In Table~\ref{tab:main_result}, we report MAE score (lower the MAE, higher the improvement) for the property prediction task. We observe that the SOTA models, when trained using \AT~generated vector representation as input, outperform their counterparts on the Materials Project and JARVIS-DFT datasets. Specifically, the average improvements of vanilla SOTA models such as CGCNN, CrysXPP, ALIGNN, Matformer, and PotNet are 21.92\%, 23.40\%, 8.63\%, 6.49\%, and 5.21\%, respectively. These improvements are significant, considering the overall architecture of these property predictor models remain unchanged, while in the input space, we introduce \AT~generated feature vectors to train these models on various downstream tasks. Additionally, it is to be noted that the average relative improvement across all properties for ALIGNN (8.63\%), Matformer (6.49\%) and PotNet (5.21\%) is lesser compared to CGCNN (21.92\%) and CrysXPP (23.40\%). The likely reason is that ALIGNN, Matformer, and PotNet are more complex models with higher parameter counts (97.8 MB for ALIGNN and 68.42 MB for Matformer) compared to CGCNN and CrysXPP. ALIGNN learns three-body interactions, Matformer captures periodic invariance, and PotNet incorporates interatomic potentials. Coversely, CGCNN and CrysXPP use simpler encoder architectures, primarily applying GCN layers to multi-graph crystal structures. Complex models often learn intrinsic features of crystal structures, so introducing atom vector representation alone doesn't significantly enhance performance. In contrast, simpler models benefit more from dense features as inputs for downstream tasks.
Another interesting observation from Table~\ref{tab:main_result} is that all SOTA models achieve an average improvement of 17.03\% in formation energy prediction, when trained using \AT generated atom vector representation. This improvement is likely because the formation energy of a crystal, defined as the difference between the energy of a unit cell composed of \( N \) chemical species and the sum of their chemical potentials (with units of $eV/atom$), depends on its node features~\cite{das2023crysgnn}. In contrast to that, However, the improvement is suboptimal for mechanical properties like bulk (9.99\%) and shear modulus (7.18\%), as these depend more on structural information such as lattice structure and symmetry~\cite{bobrowsky2018encyclopedia} than on chemical properties.
\begin{table}
  \centering
  \small
    \setlength{\tabcolsep}{6 pt}
    % \scalebox{0.75}
    % {
    \resizebox{0.55\textwidth}{!}{
      \begin{tabular}{|c |c|c c c |}
        \toprule
        Property & CGCNN & CGCNN & CGCNN & CGCNN \\
        & (\AT) & (\Skip) & (\Atom) & (\Rand)\\
        % & (Distilled) &  &   \\
        \midrule
        Formation Energy  & \textbf{0.040} (\colorbox{green!20}{-34.4}) & \ul{0.061} & 0.070 & 0.075 \\
        Bandgap (OPT)  & \textbf{0.143} (\colorbox{green!20}{-27.0})  &  \ul{0.196} & 0.251 & 0.263 \\
        Total Energy   & \textbf{0.043} (\colorbox{green!20}{-38.6})  &  \ul{0.070} & 0.076 & 0.089 \\
        Ehull & \textbf{0.124} (\colorbox{green!20}{-18.9}) &  \ul{0.153} & 0.160 & 0.164  \\
        Bandgap (MBJ) & \textbf{0.333} (\colorbox{green!20}{-20.7}) &  \ul{0.420} & 0.529 & 0.569 \\
        Bulk Modulus (Kv) & \textbf{12.37} (\colorbox{green!20}{-13.9}) & \ul{14.36}  & 15.41 & 15.99 \\
        Shear Modulus (Gv) & \textbf{10.45} (\colorbox{green!20}{-9.8}) & \ul{11.58} & 12.09 & 13.52 \\
        \bottomrule
      \end{tabular} 
      }
   \caption{\small Comparison of prediction performance (MAE) for the seven properties in JARVIS-DFT between variants of CGCNN with different dense vector representations, namely, \AT, \Skip, \Atom and \Rand. The best results have been shown in \textbf{bold} and the second best results have been \ul{underlined}. Percentage of decrease in MAE for \AT, with respect to the best performing model from the rest, has been mentioned within the bracket.}
  \label{tbl-crysatom-vs}
\end{table}
\begin{table*}[!thb]
    \centering
    \small
    \setlength{\tabcolsep}{6pt}
    % \scalebox{0.75}{
    \resizebox{1\textwidth}{!}{
    \begin{tabular}{|l |c c |c c |c c |c c |c c|}
    \toprule
	\textbf{Experiment Settings}  & CGCNN & CGCNN & CrysXPP & CrysXPP & ALIGNN & ALIGNN & Matformer & Matformer & PotNet & PotNet\\
	 &  & (\AT) &  & (\AT) &  & (\AT) &  & (\AT) &  & (\AT)\\
	\midrule
	\textbf{\vtop{\hbox{\strut Train on DFT }\hbox{\strut Test on Experimental}}} 
    & 0.265 & 0.241 (\colorbox{green!20}{-9.06}) & 0.243 & 0.222 (\colorbox{green!20}{-8.6}) & 0.220 & 0.212 (\colorbox{green!20}{-3.6}) & 0.218 & 0.213 (\colorbox{green!20}{-2.3}) & 0.217 & 0.211 (\colorbox{green!20}{-2.8})  \\
	\midrule
	\textbf{\vtop{\hbox{\strut Train on DFT and 20 \% Experimental }\hbox{\strut Test on 80 \% Experimental}}}  
	 & 0.144 & 0.111 (\colorbox{green!20}{-22.9})  & 0.138 & 0.115 (\colorbox{green!20}{-16.7}) & 0.099 &  0.093 (\colorbox{green!20}{-6.1}) & 0.098 & 0.094 (\colorbox{green!20}{-4.1}) & 0.097 & 0.093 (\colorbox{green!20}{-4.1})\\
	\midrule
	\textbf{\vtop{\hbox{\strut Train on DFT and 80 \% Experimental }\hbox{\strut Test on 20 \% Experimental}}} 
	 & 0.094 & 0.072 (\colorbox{green!20}{-23.4}) & 0.087 & 0.071 (\colorbox{green!20}{-18.4}) & 0.073  & 0.068 (\colorbox{green!20}{-6.8}) & 0.072 & 0.069 (\colorbox{green!20}{-4.2}) & 0.070 & 0.067 (\colorbox{green!20}{-4.3}) \\
    \bottomrule
    \end{tabular}
	}
	\caption{\small MAE of predicting experimental values by different SOTA models and their \AT versions with full DFT data and  different percentages of experimental data for formation energy in OQMD-EXP dataset. Percentage of decrease in MAE for \AT is mentioned in bracket.}
    \label{tab:dft_fe}
\end{table*}\\
\noindent\textbf{Comparison with Existing Distilled and Pre-trained Models.} To address the second part of RQ-1, we investigate the efficacy of utilizing a fixed atom feature vector representation in input space rather than applying resource intensive approaches such as knowledge distillation and task specific fine-tuning. As the encoder module of  CGCNN, CrysGNN and CrysXPP are variants of GCN, we have shown performance comparison between these frameworks (as shown in Table~\ref{tbl-distill-finetune}), where we have used \AT generated dense vector representation only for CGCNN. Additionally, we consider pre-trained GNN~\cite{hu2020pretraining}, which is widely used for pre-training property predictor models. We pre-train GNN~\cite{das2023crysgnn, hu2020pretraining} on 800K untagged crystal data and fine-tune it on seven different properties, as shown in Table~\ref{tbl-distill-finetune}. For fine-tuned CrysGNN framework, we consider the pre-trained encoder of CrysGNN followed by a feed-forward neural network to predict a specific property. Similarly, for distilled CGCNN framework, we apply knowledge distillation using pre-trained CrysGNN model~\cite{das2023crysgnn}. We observe that CGCNN (\AT) outperforms fine-tuned CrysGNN, CrysXPP and Pre-trained GNN with a significant margin over all properties. We also notice that CGCNN (\AT) outperforms distilled CGCNN by a large margin for formation energy, bandgap (OPT), total energy, bandgap (MBJ) and shear modulus. However, it produces comparable results for ehull and bulk modulus (Kv). The reason behind it could be use of a relatively small dataset of 139K untagged crystals to generate a fixed vector representation of atoms using \AT model, whereas the pre-training of CrysGNN is done on a large dataset of size 800K.\\
\noindent\textbf{Comparison with Existing Dense Representations.} In RQ-2, we compare performance of our dense vector representation, generated by \AT, against the existing dense representations of atoms. Here, we have considered CGCNN as the encoder module to conduct necessary experiments. We train CGCNN using our 200-dimensional atom vector representations, including \Rand\footnote{We randomly select vector representations from $\mathbb{R}^{n}$. We use normal distribution for randomly drawing the values to populate the initial vector representation.}, \Atom and \Skip on seven properties from the JARVIS-DFT dataset and reported MAE values in Table~\ref{tbl-crysatom-vs}. Our results show that CGCNN, when combined with \AT, significantly outperforms \Skip, \Atom, and \Rand aided versions across all the properties. The \Rand version performs the worst, as its dense representations fail to capture essential chemical features. The \Atom version also underperforms, due to its inability to capture the topological complexity of crystal materials. The \Skip version, which captures atomic context better than SVD-based methods, performs slightly better than \Atom but still falls short due to its limitations in comprehending complex structures.\\
\noindent\textbf{Removal of DFT Error Bias using \AT.} In this section we discuss RQ-3 to understand how we can remove DFT error bias using experimental data with the help of dense vector representations of atoms obtained from \AT. One of the fundamental issues in material science is that the experimental data instances, as described in Section~\ref{datasets}, for crystal properties are scarce~\cite{das2022crysxpp}. Hence, existing SOTA models highly rely on DFT calculated data to train its parameters. However, mathematical approximations in DFT calculation lead to erroneous prediction (error bias) in contrast to the actual experimental data~\cite{das2023crysgnn}. Hence, DFT error bias is a common problem present in the existing SOTA frameworks.  Das et.al.~\cite{das2022crysxpp} have shown that pre-training plays a significant role in mitigating error bias when fine-tuned with experimental data. Consequently, we investigate whether DFT error bias in SOTA models can be reduced with the help of our novel atom vector representation, using a small set of experimental data instances. Here, we consider OQMD-EXP~\cite{kirklin2015open} dataset to conduct the relevant experiments for formation energy prediction task. We train all SOTA models and their \AT variants with the complete DFT data in addition to a part of the experimental data. We report the MAE of different SOTA models and its \AT variants in Table~\ref{tab:dft_fe}, where the evaluation is performed on gold standard experimental data. From Table~\ref{tab:dft_fe}, we can conclude that DFT bias error in the SOTA models have reduced with the application of \AT generated vectors, consistently.
\begin{figure}[!t]%
    \centering
    % \vspace*{-1.5mm}
    {\includegraphics[width=0.60\textwidth]{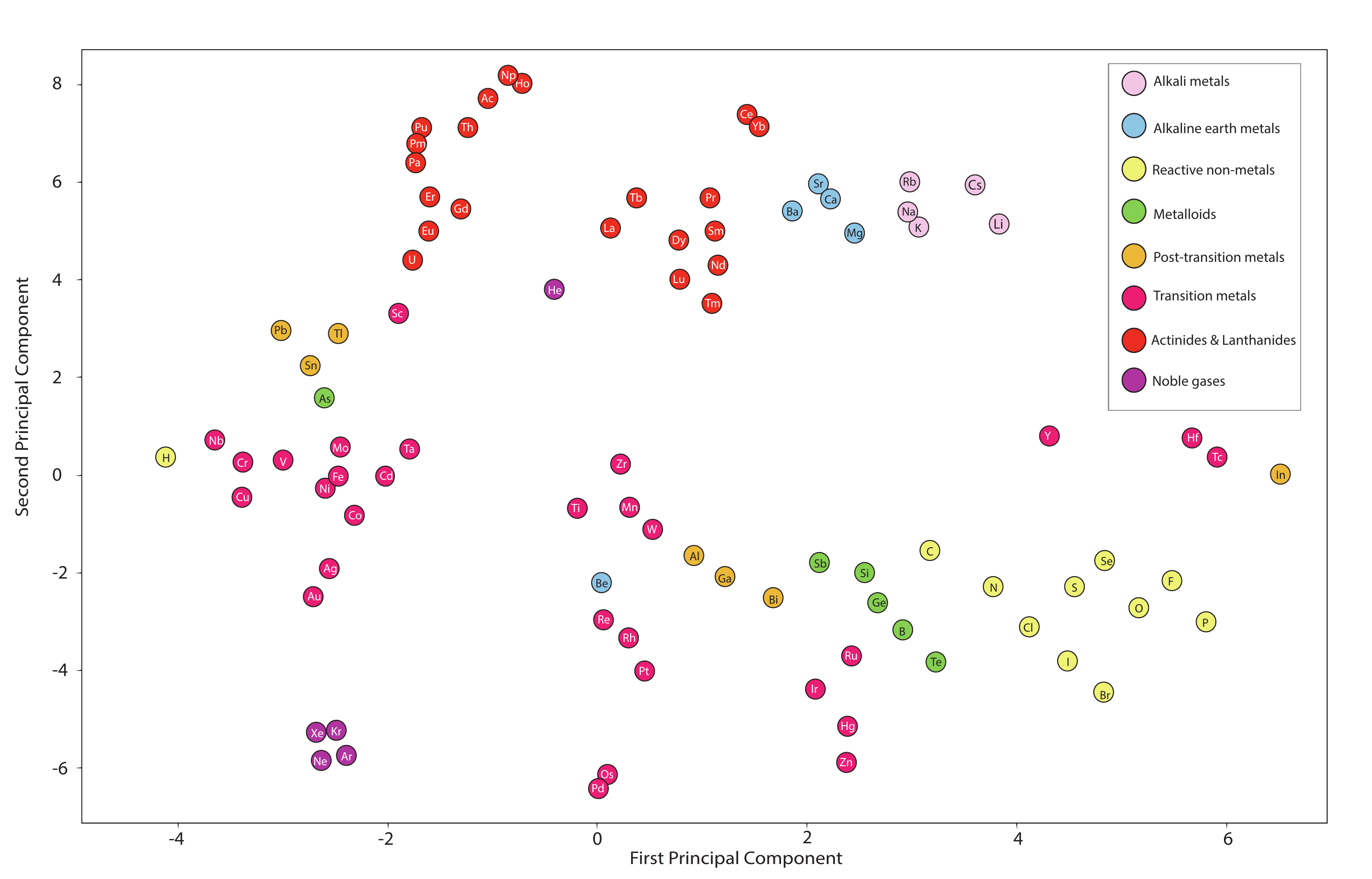}}
    \caption{\small Dimensionally-reduced atomic vectors were obtained from 200-dimensional vectors. Subsequently, these vectors were further reduced to two dimensions using t-SNE~\cite{van2008visualizing} for visualization. This plot shows an approximate position of the atomic vector representations of atoms$^{\ref{atoms-29}}$ in two dimensional Euclidean space.}
    \label{fig:atomvec}
\end{figure}\\
\noindent\textbf{Preserving Periodic Properties of Elements using \AT.} 
In this section, we focus on RQ-4 to determine if the proposed dense representations embed equivalent atomic information. To understand the chemical significance of these vector representations, we visualize them using a lower-dimensional projection\footnote{We also visualized the other two principal components namely third and fourth principal components. We observe that the projection over the third and fourth principal components is extremely noisy. There appears to be no discernible pattern beyond the first two principal components.}. The illustration in Figure~\ref{fig:atomvec} significantly aids in qualitatively examining the chemical properties of the vector representations in relation to the periodic table\footnote{We produce 200-dimensional dense representations only for 89 atoms as shown in Figure~\ref{fig:atomvec}. Remaining 29 atoms are rare elements. Consequently, our training data does not contain such atoms in compound form.\label{atoms-29}}. Our illustration shows that group-I alkali metals (Li, Na, K, Rb, Cs) form a single cluster, indicating our vector representation captures their similarity in terms of chemical properties, such as high reactivity and single valence electron~\cite{huheey2006inorganic,miessler2008inorganic,emsley2011nature,scerri2019periodic}. Similarly, group-II alkaline earth metals (Ca, Sr, Mg, Ba) cluster together, indicating their lesser reactivity and electrical conductivity. Additionally, our representation, though not illustrating Be's alkaline properties, captures its diagonal relationship with Al, indicating their tendency to form covalent bonds and tetrahedral structures~\cite{huheey2006inorganic,miessler2008inorganic,emsley2011nature,scerri2019periodic}. All reactive non-metals (C, N, O, F, Cl, Br, I, P, Se, and S) form a single cluster. This cluster spans groups VIA (halogens) and VIIA (chalcogens) of the periodic table due to their shared properties like reactivity and anion formation. Within this cluster, O and P are close in the embedding space, both binding with Hydrogen to form water and phosphine, respectively. Additionally, P and C exhibit a diagonal relationship~\cite{miessler2008inorganic,scerri2019periodic}. The figure shows that metalloid elements like Silicon (Si), Germanium (Ge), Boron (B), Antimony (Sb), and Tellurium (Te) cluster together, indicating that our vector representation captures their high semiconductivity and semi-metallic nature. Additionally, Boron and Silicon's similarity, in terms of electronegativity, allows them to form covalent bonds, showcasing their diagonal relationship. Lead (Pb), Tin (Sn), and Thallium (Tl) also form a cluster, representing post-transition metals, with Pb and Sn in close proximity due to the inert pair effect. Our representation, groups all noble gases (Ne, Ar, Kr, Xe) and captures properties of transition metals (V, Cr, Fe, Co, Ni, Cu). Lanthanides and Actinides (e.g., La, Ce, Pr, Nd, Sm, Tb, Eu, Gd, Dy, Np, Ho, Pu, Pm, Pa, U, Th, Ac, Yb, Lu, Er) cluster due to their radioactive nature. Notably, Pb and Sn also exhibit diagonal relationship properties~\cite{miessler2008inorganic, emsley2011nature, scerri2019periodic, huheey2006inorganic}.

\noindent\textbf{Investigating the Positional Validity of Hydrogen and Helium.} To investigate the second part of RQ-4, boundary cases are examined to illustrate the efficacy of the vector representation utilized in the study. Specifically, an evaluation is conducted to determine the precision of the vector representation in positioning Hydrogen (H) and Helium (He) atoms relative to their positions on the periodic table. The positioning of Hydrogen presents a contentious issue owing to its dual characteristics as both metals and non-metals, thus posing challenges in its classification within a particular metal group~\cite{rigden2003hydrogen,scerri2019periodic}. Hydrogen's electron configuration entails a single electron in its outermost shell, resembling alkali metals in group I, although it has the ability to generate a negative ion similar to halogens in group VIIA. Hence, Hydrogen is occasionally classified alongside alkali metals and at other times with halogens. Nevertheless, as depicted in Figure~\ref{fig:atomvec}, Hydrogen does not align precisely with either group, underscoring its exceptional characteristics~\cite{rigden2003hydrogen,scerri2019periodic}. Similarly, Helium, an inert gas possessing two electrons in its outer shell, diverges from the remaining noble gases in group VIIIA, and instead, it fits more appropriately within group II alongside the alkaline earth metals. The classification of Helium is further supported by Figure~\ref{fig:atomvec}, which illustrates its unique properties~\cite{scerri2019periodic}.

\section{Ablation Study}
\begin{table}[ht]
  \centering
  \small
    \setlength{\tabcolsep}{6 pt}
    % \scalebox{0.75}
    % {
    \resizebox{0.65\textwidth}{!}{
      \begin{tabular}{|c | c c c |}
        \toprule
        Property & CGCNN & CGCNN & CGCNN \\
        & (200-dim \AT) & (100-dim \AT) & (50-dim \AT) \\
        % & (Distilled) &  &   \\
        \midrule
        Formation Energy  & \textbf{0.040} & \ul{0.044} & 0.046\\
        Bandgap (OPT)  & \textbf{0.143}  &  \ul{0.160} & 0.165\\
        Total Energy   & \textbf{0.043}  &  \ul{0.049} & 0.053 \\
        Ehull & \textbf{0.124} &  \ul{0.127} & 0.130 \\
        Bandgap (MBJ) & \textbf{0.333} &  \ul{0.349} & 0.356 \\
        Bulk Modulus (Kv) & \textbf{12.37} & \ul{13.29}  & 13.60\\
        Shear Modulus (Gv) & \textbf{10.45} & \ul{10.75} & 10.99\\
        \bottomrule
      \end{tabular} 
      }
   \caption{\small Performance comparison (MAE) of various versions of CGCNN equipped with different dimensional vector representations (such as 50-dim, 100-dim, and 200-dim) obtained from \AT. The best results have been shown in \textbf{bold} and the second best results have been \ul{underlined}.}
  \label{tbl-dimensions}
\end{table}
\begin{table}[ht]
  \centering
  \small
    \setlength{\tabcolsep}{6 pt}
    % \scalebox{0.75}
    % {
    \resizebox{0.65\textwidth}{!}{
      \begin{tabular}{|c | c c c c|}
        \toprule
        Property & UL +  & UL +  & UL + & UL \\
        & SSL (BT) & SSL (VICREG) & SSL (NTXent) &  \\
        % & (Distilled) &  &   \\
        \midrule
        Formation Energy  & \textbf{0.040} & \ul{0.042} & \ul{0.042} & 0.043\\
        Bandgap (OPT)  & \textbf{0.143}  &  0.156 & \ul{0.155} & 0.158\\
        Total Energy   & \textbf{0.043}  &  \ul{0.045} & 0.047 & 0.048\\
        Ehull & \textbf{0.124} &  0.130 & \ul{0.126} & 0.127 \\
        Bandgap (MBJ) & \textbf{0.333} &  0.369 & \ul{0.340} & 0.345\\
        Bulk Modulus (Kv) & \textbf{12.37} & \ul{13.20} & \ul{13.20} & 13.24\\
        Shear Modulus (Gv) & \textbf{10.45} & \ul{10.65} & 10.68 & 10.71\\
        \bottomrule
      \end{tabular} 
      }
   \caption{\small Performance comparison of our \AT (MAE) framework by introducing different SSL loss functions such as BT, VICREG, NTXent and UL (without using SSL loss). Here we use CGCNN as our encoder model to conduct the experiments. The best results have been shown in \textbf{bold} and the second best results have been \ul{underlined}.}
  \label{tbl-UL-SSL}
\end{table}
\begin{table}[ht]
\small
% \vskip -0.1in
\begin{center}
% \vskip -0.1in
\resizebox{0.65\textwidth}{!}{
\begin{tabular}{|l|cccc|}
\toprule
Method   & Time/Epoch & Total Training Time & Total Testing Time & Model Para. \\ 
\midrule
CGCNN (\AT) & 0.189 s &  3.12 h & 0.04 s & 1.1  MB\\
CrysXPP (\AT) & 0.195 s &  3.25 h & 0.07 s & 1.1 MB  \\
ALIGNN (\AT)  & 140.4 s &  39 h & 80.4 s & 97.8 MB \\
Matformer (\AT) & 80.4 s & 22 h &  1.04 s & 68.42 MB \\
PotNet (\AT) & 42 s & 11 h & 31s & 42.9 MB\\
\bottomrule
\end{tabular}
}
\end{center}
% \vskip -0.3in
\caption{\small Training time per epoch, total training time, total testing time, and model complexity compared with CGCNN (\AT), CrysXPP (\AT), ALIGNN (\AT), Matformer (\AT) and PotNet (\AT) for formation energy on JARVIS-DFT dataset.}
\label{tab:efficiency}
\end{table}
In this section, we demonstrate the effect of variation in the dimension of the dense vector representation on its capability to encode the chemical properties of an atom. We also analyze the influence of combining UL and SSL on \AT performance and efficiency of different variants of \AT aided SOTA models by designing the following set of ablation studies:
\begin{enumerate}
\item Does increase in dimensions of the vector representation entail better performance in downstream property prediction task?
\item Does hybrid learning strategy perform well when UL is aided with SSL?
\item What is efficiency of different \AT versions of SOTA models? 
\end{enumerate}

\noindent\textbf{Impact of Increasing Dimensions.}
Results in Table~\ref{tbl-dimensions} leads to the conclusion that increase in the dimension of the dense vectors improves the performance of the property predictor model (here, CGCNN). Our conclusion is aligned with earlier findings by Antunes et..al~\cite{antunes2022distributed}. However, increasing dimension size of the vector representation beyond a certain point may lead to exponential growth in computational resources and may prove to be an impediment in terms of model training. 
%increasing dimensions of the distributed representation of atoms captures more information in it and produces improvement in result this claim was clearly observed in our result that increasing the dimension of our generic distributed representations of atoms lower's the MAE for all the properties. We note that a 200-dimensional atom vector performs better than 100-dimensional and 50-dimensional vectors for all seven properties. This result indicates that higher dimensional vector representation improves the performance of the baseline property predictor models significantly. However, increasing dimension size of the vector representation may lead to exponential growth in computational resources and may prove to be an impediment in terms of model training.

\noindent\textbf{Impact of Combining UL and SSL.} Table~\ref{tbl-UL-SSL} demonstrates how effectively we can combine our UL and SSL based approaches in JARVIS-DFT dataset for seven properties. We observe that combination of SSL and UL based frameworks equipped with Barlow Twins~\cite{zbontar2021barlow} loss helps the existing property predictors significantly to achieve SOTA performance on downstream task, when compared with NTXent~\cite{chen2020simple} and VICREG~\cite{bardes2021vicreg} losses. The common intuition is that, while training a model using Barlow Twins loss, it generates the hard negative samples by masking the nodes and edges internally for a graph to capture the overall topological structure. Whereas, SSL loss functions such as NTXent, VICREG take necessary negative samples by considering external properties such as space group information. SSL strategies, in isolation, are not used for generative tasks, thereby preventing us from using such strategies for generation of dense vector representations. However, UL is applied together with SSL, leading to a generative loss for generating a dense vector representation as a subtask of graph generation.

\noindent\textbf{Efficiency of \AT Variant of the SOTA Models.} Table~\ref{tab:efficiency} shows comparison of CGCNN (\AT), CrysXPP (\AT), ALIGNN (\AT), Matformer (\AT) and PotNet (\AT) in terms of training time per epoch, total training time, total testing time and model complexity for formation energy on JARVIS-DFT dataset. We clearly observe that CGCNN (\AT) is the fastest model among all the \AT variant of SOTA models. It uses less number of parameters compared to the other models. This led us to consider CGCNN as our encoder model to conduct the necessary experiments in ablation study.  

% In this work, we present a novel framework \AT to generate dense representation vectors of atoms that can be used in various graph neural network-based downstream property prediction tasks. In our framework, we have combined both UL and SSL techniques to generate dense representations of atoms. In UL, we employ auto-encoder decoder based framework with reconstruction loss function and we utilize encoder based SSL strategy by utilizing Barlow Twins loss function. We use our dense vectors in various downstream property prediction tasks like prediction of formation energy, bandgap, etc., and we also generate vector representations of molecules from atoms to establish the chemical equivalence of our generated dense vectors. We use t-SNE algorithm to visualize our dense representation of atoms to establish chemical validity. An interesting extension of this work could be to generate a generalized vector representation of atoms that can be used in organic chemistry, along-with a representation of bonds present in the organic molecules. Another extension of our work is by incorporating many body interactions in \AT to produce better average improvement in the complex SOTA models.
\section{Conclusion}
In this study, we introduce a novel framework, \AT, designed to create dense vector representations for crystal atoms. These vectors play a pivotal role for different graph neural network-based crystal property prediction tasks.  Our approach uniquely combines Unsupervised Learning (UL) and Self-Supervised Learning (SSL) techniques to generate these dense representations. Additionally, we propose a novel way to extract the generalized feature vector representation from the latent space of the encoder module of \AT framework. Our empirical results demonstrate that \AT  significantly enhances the performance of existing neural property predictors. Experiments show that our proposed framework generates a robust and unbiased dense vector representation for atoms, effectively capturing periodic properties and chemical significance of atoms. Future directions for extending our work could be incorporating many-body interactions as a part of the system, aiming to achieve performance improvements across complex state-of-the-art models.

% For the UL component, we employ an auto-encoder-decoder framework, optimizing it with a reconstruction loss function. Concurrently, the SSL strategy leverages the Barlow Twins loss function within an encoder setup.

%  To validate the chemical relevance of our representations, we utilize the t-SNE algorithm for visualization. This visualization confirms the effectiveness of our dense atomic representations in capturing chemical properties. 

%
% For natbib users:
\bibliographystyle{unsrtnat}
\bibliography{reference}
% For bibLaTeX users:
% \printbibliography

% \appendix
% \section{Appendix}
% Any possible appendices should be placed after bibliographies.
% If your paper has appendices, please submit the appendices together with the main body of the paper.
% There will be no separate supplementary material submission.
% The main text should be self-contained; reviewers are not obliged to look at the appendices when writing their review comments.

\end{document}